# GeV Gamma-Rays from Molecular Clouds Illuminated by Particles Diffusing from the Adjacent Supernova Remnant G335.2+0.1 Confined in an Expanding Bubble

Chen Huang[1] , Xiao Zhang[2] , Yang Chen[1,3] , Qian-Qian Zhang[1] , Wen-Juan Zhong[1] , and Xin Zhou[4]
[1] School of Astronomy & Space Science, Nanjing University, 163 Xianlin Avenue, Nanjing 210023, People's Republic of China; ygchen@nju.edu.cn
[2] School of Physics & Technology, Nanjing Normal University, No.1 Wenyuan Road, Nanjing 210023, People's Republic of China; xiaozhang@njnu.edu.cn
[3] Key Laboratory of Modern Astronomy and Astrophysics, Nanjing University, Ministry of Education, Nanjing 210023, People's Republic of China
[4] Purple Mountain Observatory and Key Laboratory of Radio Astronomy, Chinese Academy of Sciences, 10 Yuanhua Road, Nanjing 210023, People's Republic of China
*Received 2025 May 1; revised 2025 July 23; accepted 2025 July 24; published 2025 September 10*

## Abstract

We report the detection of GeV gamma-ray emission likely associated with supernova remnant (SNR) G335.2+0.1 and the finding of a molecular cloud ($\sim 20'$–$30'$ in angular size) that is very likely in physical contact with the SNR and responsible for the gamma-ray emission. Using the 16.8 yr Fermi-LAT data, an extended emission, with a significance of $13.5\sigma$ and a radius $0°.24$ in 0.2–500 GeV in the uniform-disk model, was found to the adjacent east of the SNR. With archival Mopra CO-line data, a large molecular clump at local-standard-of-rest velocity $\sim -48$ to $-43$ km s$^{-1}$ was revealed, appearing coincident with the gamma-ray source. The SNR was found located in a cavity encircled by a "C"-shaped ring-like molecular shell at $-45$ to $-43$ km s$^{-1}$. This morphological agreement, together with the position-velocity diagrams made along lines cutting across the cavity, suggests that the SNR was evolving in the expanding molecular bubble created by the stellar wind of the progenitor with a mass $\gtrsim 15~M_\odot$. The giant molecular cloud, visible at around $-46$ km s$^{-1}$, and the associated SNR are thus estimated to lie at a kinematic distance of 3.1 kpc, with the H I absorption taken into account. We suggest that the SNR has entered the radiative phase after the blastwave recently struck the cavity wall. With the evolutionary scenario of the SNR, we demonstrate that the gamma-ray emission reported here can be naturally interpreted by the hadronic interaction between the accelerated protons escaped from the SNR shock and the eastern large molecular clump.

*Unified Astronomy Thesaurus concepts:* Gamma-rays (637); Supernova remnants (1667); Molecular clouds (1072); Cosmic rays (329)

## 1. Introduction

As one of the most energetic sources and a prime candidate for accelerators of Galactic cosmic rays (CRs), supernova remnants (SNRs) are capable of accelerating particles to relativistic velocity via diffusive shock acceleration (e.g., L. O. Drury 1983; M. A. Malkov & L. O. Drury 2001; K. M. Schure & A. R. Bell 2013). The accelerated protons can produce gamma-ray emission through neutral-pion decay following proton–proton collisions with dense material. Given that protons dominate the CR population, detecting their acceleration signatures via SNR-associated hadronic gamma-ray emission provides critical evidence for understanding the particle acceleration and transport mechanisms.

Radio source G335.2+0.1 was first discovered by D. H. Clark et al. (1973) and later identified as a filamentary shell SNR ($\sim 21'$ in diameter) with radio data from the south Galactic survey (J. B. Z. Whiteoak & A. J. Green 1996). P. Eger et al. (2011) continued a corresponding multi-wavelength study and found a morphological match in the southwestern part of the SNR with a clump at local-standard-of-rest (LSR) velocity $-27$ to $-18$ km s$^{-1}$ using $^{12}$CO data from the Nanten Galactic plane survey. This cloud overlaps projectively with the TeV source HESS J1626−490, discovered by the High Energy Stereoscopic System (H.E.S.S., F. Aharonian et al. 2008a). With the XMM-Newton EPIC observation, several X-ray point sources were found at the location of the TeV source, but unrelated to the detected TeV gamma-ray emission (P. Eger et al. 2011). Within the SNR, the PSR J1627−4845 is offset from the nominal remnant center by $\sim 4'$. This pulsar, with a characteristic age of $\tau = 2.7 \times 10^6$ yr, is much older than the typical maximum age assumed for an observable SNR and suggested to be a chance alignment (V. M. Kaspi et al. 1996).

Due to the possible connection between the TeV source HESS J1626−490 and SNR G335.2+0.1, we aimed to search for the potential GeV counterpart by analyzing Fermi data. Theoretically, investigating GeV–TeV gamma-ray emission associated with SNRs could contribute to the study of the shock acceleration mechanism and the transport of escaped particles. However, associating SNRs and gamma-ray emission is sometimes not very easy. Multiwavelength investigations are required to search for the potential spatial coherence across different energy bands and to constrain the SNR parameters. In particular, analysis of the surrounding molecular environment is crucial for revealing the gamma-ray radiation mechanisms.

In this work, we analyze the Fermi-LAT gamma-ray data toward the SNR G335.2+0.1 and its surrounding interstellar environment. In Section 2, we describe the details of the data used here. Analysis of the observational data and the corresponding results are presented in Section 3. Then, in Section 4, we discuss the distance and dynamical evolution of the SNR and present a hadronic interpretation for the observed gamma-ray emission. Finally, we conclude this study in Section 5.

## 2. Observations and Data

### 2.1. Fermi-LAT Observational Data

We analyzed more than 16.8 yr (from 2008 August 4 15:43:36 (UTC) to 2025 May 30 05:21:22 (UTC)) of Fermi-LAT Pass 8 SOURCE class (evclass = 128, evtype = 3) data with the software Fermitools 2.2.0.[5] The regions of interest (ROIs) in our study are $15° \times 15°$ in size, centered at the position of SNR G335.2+0.1 (R.A.$_{J2000}$ = 246°.91, decl.$_{J2000}$ = −48°.74).

First, the data selection was made with the command *gtselect* with the maximum zenith angle of 90° to reduce the contamination from the Earth limb. Then, we applied the command *gtmktime* to the data with the recommended filter string "(DATA_QUAL >0)&&(LAT_CONFIG == 1)" for choosing good time intervals. The entire energy range from 0.2 to 500 GeV was divided into 10 logarithmic bins per decade for the counts cube and the exposure cube. The appropriate Instrument Response Functions are "P8R3_SOURCE_V3." The Galactic interstellar diffuse background emission model "gll_iem_v07" and isotropic background spectral template "iso_P8R3_SOURCE_V3_v1," as well as the sources listed in the Fermi-LAT 14 yr source catalog (4FGL-DR4, J. Ballet et al. 2023) within a radius of 25° from the ROI center, were incorporated for analysis using the user-contributed tool *make4FGLxml.py*.[6] After that, we used the Python module *pyLikelihood* with the NEWMINUIT optimizer to perform the binned likelihood analysis and get the best-fit results. In this step, we only freed spectral parameters of the catalog sources within 5° of the ROI centers and the normalization of the two diffuse background components. In addition, the Python package Fermipy (M. Wood et al. 2017) was employed[7] (version 1.2) in the position, extension, and spectral energy distribution (SED) fitting process.

### 2.2. CO Observations

We analyzed the $^{12}$CO $J = 1\text{–}0$ (115.271 GHz) and $^{13}$CO $J = 1\text{–}0$ (110.201 GHz) data from the Mopra Carbon Monoxide (CO) Survey of the Southern Galactic Plane—Data Release 3 (M. G. Burton et al. 2013; C. Braiding et al. 2018). The mapped region is centered at the Galactic coordinates $l = 335°.00$, $b = 0°.00$ and has a size of $\sim 1° \times 1°$ covering the G335.2+0.1. These data have been taken at 0'.6 spatial resolution and 0.1 km s$^{-1}$ spectral resolution. The main-beam brightness temperature $T_{\rm mb}$ was obtained by dividing the given antenna temperature $T_{\rm A}^*$ by the extended beam efficiency $\eta_{\rm XB} = 0.55$. The average rms noise (defined in $T_{\rm mb}$) is $\sim 2.8$ K in the $^{12}$CO data and $\sim 1.4$ K in the $^{13}$CO data.

### 2.3. Other Data

We also used the new radio continuum data at 1.3 GHz from SARAO MeerKAT Galactic Plane Survey (SMGPS; S. Goedhart et al. 2024) and H I as well as 1.4 GHz continuum data from the Southern Galactic Plane Survey (SGPS;

---

[5] https://fermi.gsfc.nasa.gov/ssc/data/analysis/software/
[6] https://fermi.gsfc.nasa.gov/ssc/data/analysis/user/make4FGLxml.py
[7] https://fermipy.readthedocs.io/en/latest/

N. M. McClure-Griffiths et al. 2005; M. Haverkorn et al. 2006). The radio continuum image has a spatial resolution of $\sim 8''$ and a sensitivity of $\sim 10\text{–}20$ $\mu$Jy beam$^{-1}$. As for the H I line, the spatial resolution and spectral resolution are $\sim 2'$ and 0.8 km s$^{-1}$, respectively, with the rms noise sensitivity of $\sim 1.6$ K.

## 3. Data Analysis and Results

### 3.1. Fermi-LAT Data Analysis

#### 3.1.1. Spatial Analysis

We first generated a test-statistic (TS) map in the energy range 1–500 GeV (see the left panel of Figure 1) utilizing the command *gttsmap* by including 4FGL-DR4 catalog sources and two diffuse background components (detailed in Section 2.1) as models to check the residual emission around our target. The TS value for each pixel was evaluated by TS = $-2\ln(\mathcal{L}_0/\mathcal{L}_1)$, where $\mathcal{L}_0$ is the maximum likelihood of the null hypothesis and $\mathcal{L}_1$ is the maximum likelihood of the test model that a putative point source with a fixed index of 2 was located in this pixel. As can be seen, the catalog source 4FGL J1628.2−4848c (hereafter called J1628) is within the SNR shell and regarded as a possible GeV counterpart of SNR G335.2+0.1. We found that there is some residual excess around the SNR and J1628 and thus generated a TS map (see the right panel of Figure 1) by excluding source J1628 from the source model. As shown in the figure, a residual emission in the green circle partially overlaps with the SNR, with the peak TS located in close proximity to the east of the SNR shell.

Since the catalog source J1628 failed to fit the residual excess, three spatial models of a point source, a disk, and a Gaussian model were separately used to refit the data in the energy range of 1–500 GeV, for which the extensions and locations were optimized via the *extension* and *localize* methods in Fermipy to maximize the likelihood. Meanwhile, the significance of extension was evaluated by comparing the likelihood of an extended source hypothesis with that of a point-source hypothesis. The gamma-ray source is considered to be significantly extended if TS$_{\rm ext} = 2\ln(\mathcal{L}_{\rm ext}/\mathcal{L}_{\rm ps}) \geqslant 16$ (J. Lande et al. 2012). For all three models, the spectral types were postulated to be LogParabola (LogP) spectrum (that is, $dN/dE = N_0 (E/E_0)^{-\Gamma - \beta \ln(E/E_0)}$) with $E_0$ fixed to be 2658 MeV as that for the catalog source J1628. The point-source model was fitted by directly reoptimizing the location and spectral parameters of the source.

Table 1 shows the fitted parameters for the three hypotheses. To choose the best proper model, the Akaike information criterion (AIC, H. Akaike 1974) was adopted for each case. The AIC is defined as AIC = $2k - 2\ln\mathcal{L}$, where $k$ is the number of free parameters in the model and $\mathcal{L}$ is the maximum likelihood estimate. By comparing AIC between different models, the one with the minimum AIC performs better than the others. It can be seen that the disk and Gaussian hypotheses present a remarkable improvement compared with the point-source hypothesis. Moreover, the obtained TS$_{\rm ext}$ values 54.15 and 52.27 for the disk and Gaussian models, respectively, also mean that an extended source can describe the emission more properly beyond a point source, and the disk model is more significant than the Gaussian model. The centroids in the disk, Gaussian, and optimized J1628 models are consistent with each other in the range of 1$\sigma$ positional uncertainty after considering systematic error.[8] The spectral parameters included in Table 1 show harder spectra (smaller photon indices) and reasonably larger normalization for the cases of extended sources than for the case of a point source.

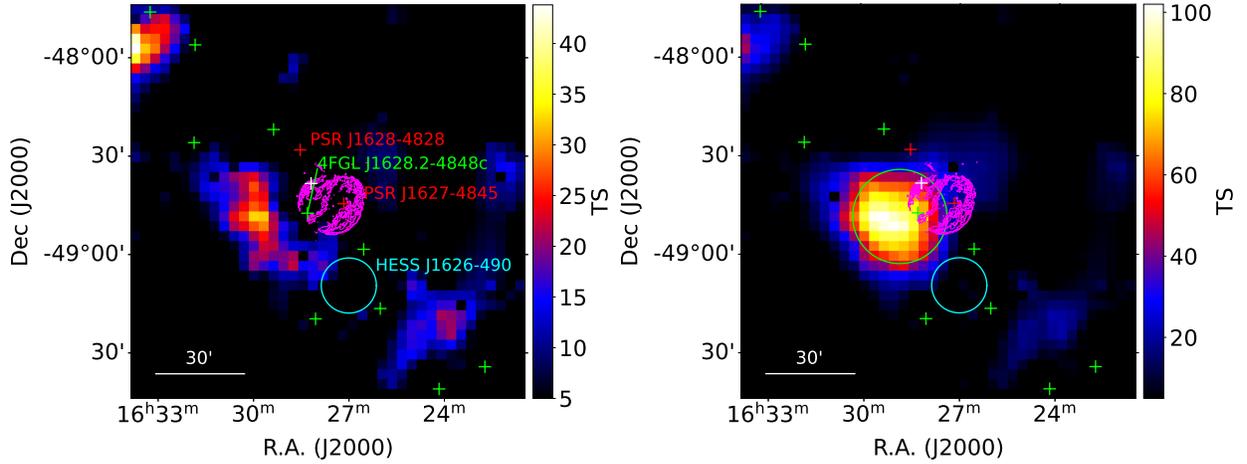

**Figure 1.** TS maps of 2° × 2° regions centered at SNR G335.2+0.1 in the energy range 1–500 GeV for an improved angular resolution. Left panel: the TS map with the 4FGL-DR4 catalog sources model. Right panel: the TS map using the same model, but with source 4FGL J1628.2−4848c excluded. The green pluses mark the positions of 4FGL-DR4 catalog sources (J. Ballet et al. 2023). The cyan circle indicates the 1σ Gaussian radius of HESS J1626−490 (H. E. S. S. Collaboration et al. 2018). The magenta contours represent 0.4, 0.8, and 1.2 mJy beam$^{-1}$ levels of the radio continuum emission of the SNR from MeerKAT observations at 1359.7 MHz (S. Goedhart et al. 2024). The green circle represents 68% extension of the best-fitted spatial model (disk). The two pulsars within 0°.5 radius of the extended source center are depicted in red pluses. The white plus (IGR J16283−4838) marks the position of the High Mass X-ray Binary within the 68% extension of the disk model.

### 3.1.2. Spectral Analysis

Once the spatial analysis was completed, we applied the obtained disk model in the spectral analysis over the whole energy range of 0.2–500 GeV. During the fitting, the spectral parameters of sources located within 5° of the ROI center vary freely, as well as the normalization of the two background models. To study the spectral properties of the disk, in addition to LogP, we tested three spectral types: PowerLaw (PL), ExpCutoffPowerLaw (ECPL), and BrokenPowerLaw (BPL) in separate fits. The formulas and ΔAIC values for different photon distributions are shown in Table 2. Considering the large gap in ΔAIC between spectral types PL/BPL and ECPL/LogP, ECPL and LogP apparently have high probabilities, with ECPL slightly better. For ECPL, with $E_0 = 2658$ MeV, the best-fit spectral values are $N_0 = (2.83 \pm 0.33_{stat} \pm 0.85_{sys}) \times 10^{-12}$ MeV$^{-1}$ cm$^{-2}$ s$^{-1}$, $\Gamma = 0.11 \pm 0.20$, and $E_{cut} = 1881 \pm 125$ MeV. For LogP, the best-fit spectral values are $N_0 = (6.83 \pm 0.54_{stat} \pm 1.78_{sys}) \times 10^{-13}$ MeV$^{-1}$ cm$^{-2}$ s$^{-1}$, $\Gamma = 1.81 \pm 0.15$, and $\beta = 0.68 \pm 0.13$. The luminosity between 0.2 and 500 GeV is $1.8 \times 10^{34} (d/3.1\,\text{kpc})^2$ erg s$^{-1}$ and $1.9 \times 10^{34} (d/3.1\,\text{kpc})^2$ erg s$^{-1}$ for ECPL and LogP, respectively, with a reference value 3.1 kpc adopted for distance $d$ to the emission source (see Section 4.1). Because of the tiny discrepancy of ΔAIC between ECPL and LogP, the spectral type of the source is difficult to characterize in the LAT energy range, and for simplicity, we still adopt the spectral model LogP for this disk model as above. The TS value of the disk source is 205, corresponding to the significance of 13.5σ with 6 degrees of freedom.

Based on the maximum likelihood analysis, the SED of this source was produced by the *sed* method in Fermipy in nine logarithmically spaced energy bins, as shown in Figure 2. Due to the low significance, the last four bins were combined into one bin to make a global limit. During the fitting process, the free parameters only include the normalization of the sources with significance ⩾5σ and within 3° from the ROI centers, as well as the Galactic and isotropic diffuse background components, while all of the other parameters were fixed to their best-fit values in the global fitting. For the energy bins with TS ⩽ 4, the 99%-confidence-level upper limits were calculated.

In addition to statistical errors, the systematic uncertainty of the spectrum from the choice of the Galactic interstellar emission model was simply estimated by artificially varying the normalization of the Galactic diffusion model by ±6% from the best-fit values in the entire energy band as well as in the individual energy bins of SED (A. A. Abdo et al. 2009). Then, the maximum deviations of flux between normal fitting and the fitting with changed normalization of the Galactic diffusion model were considered as the systematic errors. These two kinds of errors were combined in quadrature.

### 3.1.3. Temporal Variation Analysis

In order to examine the long-term variability of this gamma-ray source, a 3 month binned light curve was constructed (see Figure 3) over the total time span and the 0.2–500 GeV energy range by using the *lightcurve* method in Fermipy. The bin size could be flexible from several days to several months, and 3 months was adopted for convenience. Using the criterion in P. L. Nolan et al. (2012), for 68 time bins, the variability index (TS$_{var}$) was required in excess of 98.0 to identify a 99% temporal variability. In view of the obtained TS$_{var}$ = 51.8, no significant gamma-ray variability can be counted as being detected.

## 3.2. Properties of the Molecular Gas

### 3.2.1. Molecular Cavity/Shell Around −46 km s$^{-1}$

In view of the detected gamma-ray excess that is probably associated with SNR G335.2+0.1, we next explored the dense

---
[8] The total uncertainty radius was calculated as $r_{tot}^2 = (1.06 r_{stat})^2 + r_{abs}^2$, where $r_{abs} = 0°.0068$ (S. Abdollahi et al. 2020).

Table 1
Results of the Spatial Analysis of the GeV Emission Around G335.2+0.1 in 1–500 GeV

| Model | R.A., decl. (°) | Extension (°) | Spectral Parameters | ΔAIC |
|---|---|---|---|---|
| 4FGL J1628.2−4848c (point source) | $247.30 \pm 0.°03, -48.83 \pm 0.°02$ | ⋯ | $N_0 = 4.52 \pm 0.32, \Gamma = 2.22 \pm 0.14, \beta = 0.83 \pm 0.16$ | 0 |
| Disk | $247.21 \pm 0.°02, -48.81 \pm 0.°03$ | $0.24^{+0.02}_{-0.02}$ ° | $N_0 = 7.63 \pm 0.68, \Gamma = 1.97 \pm 0.20, \beta = 0.64 \pm 0.19$ | −53.5 |
| Gaussian | $247.20 \pm 0.°03, -48.80 \pm 0.°03$ | $0.27^{+0.04}_{-0.03}$ ° | $N_0 = 8.47 \pm 0.70, \Gamma = 1.99 \pm 0.17, \beta = 0.64 \pm 0.17$ | −53.3 |

**Note.** The values of $N_0$ are given in units of $10^{-13}$ MeV$^{-1}$ cm$^{-2}$ s$^{-1}$. The extension for the disk and Gaussian models refers to their respective 68%-containment radii. ΔAIC was defined as the difference between the AIC value for the model with the lowest AIC value and that of the point-source model. The $1\sigma$ statistical uncertainties were given for morphological and spectral parameters. All parameters in this table were determined in the analysis of events with energies above 1 GeV.

environmental gas around it, based on CO line observation, for the possibility of hadronic interaction. The average main-beam brightness temperature of $^{12}$CO emission toward the SNR region (within 11′ from the center) shows multiple peaks at the LSR velocities $V_{\rm LSR} \sim -113, -94, -86, -70, -46, -40$, and $-27$ km s$^{-1}$ (see Figure 4). Only components at around $-86$ and $-46$ km s$^{-1}$ have a spatial correspondence with the SNR and the extended GeV gamma-ray emission mentioned above. In the velocity range $\sim -90$ to $-80$ km s$^{-1}$, neither cavity-like and filamentary structures with spatial correspondence to the SNR nor broadened CO line profiles along the SNR boundary are found, and therefore the molecular gas in this velocity range is regarded as irrelevant to the SNR. At around $-46$ km s$^{-1}$, spatial distribution of the $^{12}$CO emission is shown in Figure 5 with velocity intervals of 0.7 km s$^{-1}$. It is seen that a molecular cloud $\sim 20'$–$30'$ in size is adjacent to, and projectively overlaps with, the SNR. Moreover, in the velocity interval $-45$ to $43$ km s$^{-1}$, a "C"-shaped molecular ringlike structure encircles the SNR in the north, east, and south. The SNR thus appears to be confined in a molecular gas cavity by an incomplete molecular shell. Also, a large molecular clump, named hereafter as "Region G," to the east of the SNR at a velocity from $-48$ to $-43$ km s$^{-1}$ appears coincident with the extended GeV gamma-ray emission discussed in Section 3.1. The GeV gamma-ray emission, the $^{12}$CO ($J = 1$–0) line emission, and the radio emission of the SNR are shown together in Figure 6 for morphological comparison.

### 3.2.2. Kinematic Signature of the Expanding Molecular Shell

In order to search for broadened line profiles of CO emission around $-46$ km s$^{-1}$ as a kinematic signature of the SNR and molecular cloud (MC) interaction, we checked a $^{12}$CO line profile grid toward the northeastern and southern boundary, where the radio emission is bright. Due to complex line crowding along the line of sight and large rms noise, broad intermixed line profiles span from $-50$ to $-35$ km s$^{-1}$ and line broadening due to shock disturbance cannot be discerned.

To further test whether the cavity confined by the molecular ringlike structure or shell revealed around $-46$ km s$^{-1}$ is physically associated with SNR G335.2+0.1, we then inspected the position–velocity (PV) diagrams (shown in Figure 7) along lines crossing through the geometric center of the SNR (R.A.$_{J2000} = 246.°87$, decl.$_{J2000} = -48.°75$) at different positional angles (see Figure 5). All of the PV diagrams display arclike or elliptical structures across the remnant between velocity $-47$ to $-37$ km s$^{-1}$. Such an ellipse pattern in the PV diagram typically represents expansion motion of molecular gas surrounding the SNR due to the Doppler effect. As will be discussed in Section 4.2, the expansion may be driven by the stellar wind of the progenitor star of the SNR. Therefore, the PV curve pattern provides a kinematic signature for the association of the MC with the SNR and its progenitor.

In addition, among the upper panels of Figure 7, several possible broadened structures (marked by green boxes) were revealed by $^{12}$CO emission with a velocity width of $\sim 8$ km s$^{-1}$, which could be attributed to a high-speed outward motion of a part of the molecular gas.

### 3.2.3. Parameters of the Associated Molecular Gas

With the association established between the SNR and the molecular gas around $-46$ km s$^{-1}$, we estimate some basic parameters of the environmental molecular gas in two regions, "Region G" and "Region W" (delineated in Figure 5), over the velocity ranges $-48$ to $-43$ km s$^{-1}$ and $-45$ to $-43$ km s$^{-1}$, respectively. Assuming local thermodynamic equilibrium and that $^{12}$CO and $^{13}$CO lines are optically thick and optically thin, respectively, the H$_2$ column density is obtained via $N(\text{H}_2) = 1.49 \times 10^{20} W(^{13}\text{CO})/[1 - \exp(-5.29/T_{\rm ex})]$ cm$^{-2}$ (T. Nagahama et al. 1998), where $W(^{13}\text{CO})$ is the average integrated intensity of $^{13}$CO line emission and $T_{\rm ex}$ is the excitation temperature derived from $^{12}$CO line as $T_{\rm ex} = 5.53[\ln(1 + 5.53(T_{\rm mb}/f + 0.836)^{-1})]^{-1} \sim 13.4$ K, where the filling factor $f$ was adopted as 0.8. Thus, the mass of gas is obtained from $M = \mu m_{\rm H} N(\text{H}_2) A$ with mean atomic weight $\mu = 2.8$ and $A$ the cross-sectional areas of "Region G" and "Region W." As the spatial distribution of the gas could be complex and a part of the gas (in the assumed LSR velocity range adopted) may be unrelated to this associated system but instead projected along the line of sight, the molecular gas mass could be overestimated and should be taken with caution.

For Region G and Region W, the molecular gas densities are estimated using line-of-sight depths similar to the transverse sizes and adopting a distance of 3.1 kpc to the SNR–MC association (Section 4.1). These parameters of the molecular gas are listed in Table 3. According to the spatial distribution of optically thin $^{13}$CO emission (shown in Figure 5), the distances from the near and far edges of "Region G" to the SNR center are $L_1 \approx 11$ pc and $L_2 \approx 20$ pc, respectively, which will be used in the discussion of the gamma-ray emission below (Section 4.4).

**Table 2**
Formulae and Likelihood Test Results for Gamma-Ray Spectra in 0.2–500 GeV

| Name | Formula | Free Parameters | Flux ($\times 10^{-11}$ erg cm$^{-2}$ s$^{-1}$) | $\Delta$AIC |
|---|---|---|---|---|
| PL | $dN/dE = N_0(E/E_0)^{-\Gamma}$ | $N_0, \Gamma$ | 3.3 | 0 |
| ECPL | $dN/dE = N_0(E/E_0)^{-\Gamma}\exp(-E/E_{\rm cut})$ | $N_0, \Gamma, E_{\rm cut}$ | 1.6 | $-31.9$ |
| LogP | $dN/dE = N_0(E/E_0)^{-\Gamma - \beta\ln(E/E_0)}$ | $N_0, \Gamma, \beta$ | 1.7 | $-27.9$ |
| BPL | $dN/dE = N_0 \begin{cases}(E/E_b)^{-\Gamma_1}, E \leqslant E_b \\ (E/E_b)^{-\Gamma_2}, E \geqslant E_b\end{cases}$ | $N_0, E_b, \Gamma_1, \Gamma_2$ | 2.7 | $-9.7$ |

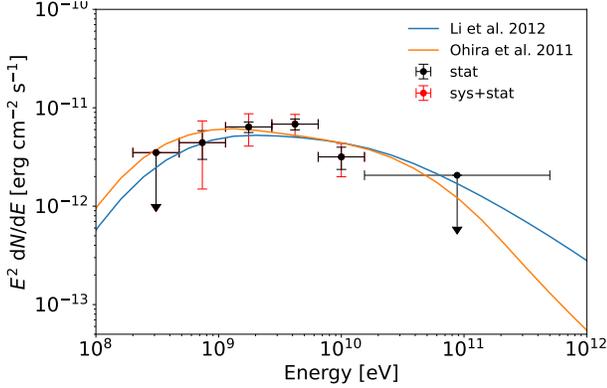

**Figure 2.** Spectral energy distribution of the SNR G335.2+0.1. The black and red points represent gamma-ray flux with only statistical uncertainty considered and the combination of statistical and systematic uncertainty, respectively. The model curve in blue is produced with the diffusion model of H. Li & Y. Chen (2012). The curve in orange is produced utilizing the $\delta$-escape diffusion model of Y. Ohira et al. (2011).

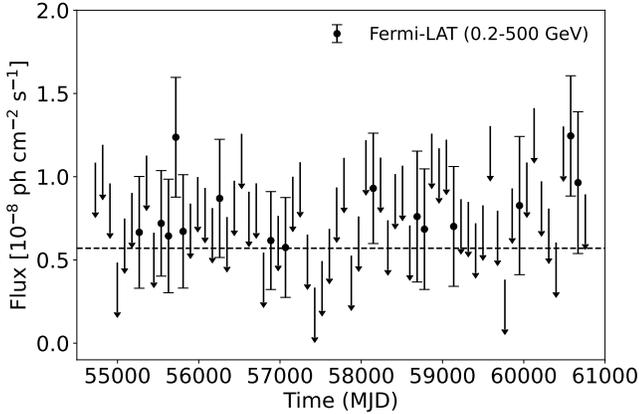

**Figure 3.** The 3 month binned gamma-ray light curve in the disk model. The horizontal dashed line represents the constant flux for the LogP model (see Table 2). For the bins with TS $\leqslant 4$, the 95% upper limits are presented.

## 4. Discussion

### 4.1. Distance to the Supernova Remnant–Molecular Cloud Association

An estimate of the kinematic distance, 1.8 kpc, to the SNR G335.2+0.1 was provided based on both the CO data and the brightness temperature depression in the H I gas at $V_{\rm LSR} = -22.5$ km s$^{-1}$ (P. Eger et al. 2011; S. Ranasinghe & D. Leahy 2022). Alternative distance estimates put it at $\sim$4 kpc according to the surface-brightness-to-diameter relations (M. Z. Pavlović et al. 2013) and near-infrared extinction (S. Wang et al. 2020).

However, the SNR–MC association at around $-42$ km s$^{-1}$, which is revealed in this work by morphological agreement between the molecular cavity and SNR together with the expanding motion of the molecular shell confining the SNR (Section 3.2), can independently pinpoint the kinematic distance to the SNR.

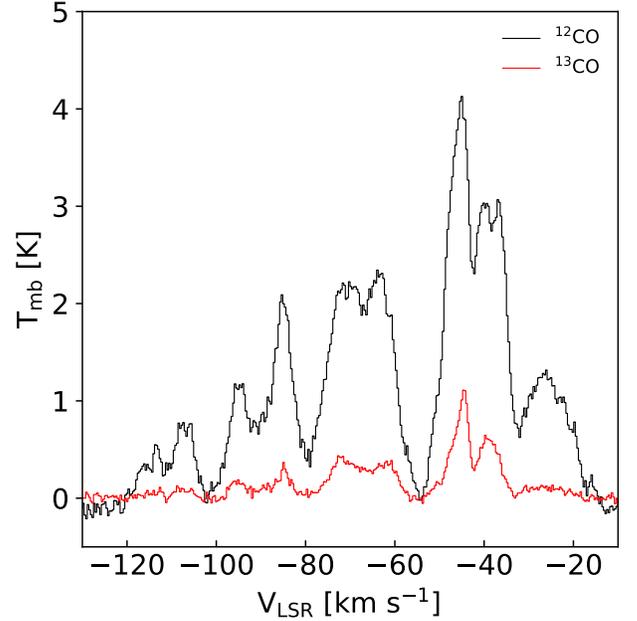

**Figure 4.** Spectra of $^{12}$CO $J = 1$–0 and $^{13}$CO $J = 1$–0 line emission in the SNR region.

By comparison, in the $^{12}$CO channel map (Figure 5), there is no obvious morphological sign of SNR–MC association at around $-22.5$ km s$^{-1}$.

Adopting the traditional Galactic rotation model (J. Brand & L. Blitz 1993) and updated Solar parameters $R_0 = 8.34$ kpc (distance from the Sun to the Galactic center) and $\Theta_0 = 40$ km s$^{-1}$ (circular rotation speed at the position of the Sun) (M. J. Reid et al. 2014), the distances for the systemic peak velocity of the associated MC $V_{\rm LSR} \sim -46$ km s$^{-1}$ which obtained in Figure 4 are 3.1 kpc (near) and 12.0 kpc (far).

In order to obtain further constraints on the distance estimates, we studied the H I absorption feature toward the SNR portions that are bright in radio continuum. The continuum-subtracted H I observations were extracted from the SGPS. However, we cannot identify clear absorption features due to the large rms ($\sim$1.6 K), and then the integrated H I maps were plotted. As can be seen in Figure 8, there is a deep absorption cavity located at around $-22.5$ km s$^{-1}$, which agrees with the results in P. Eger et al. (2011). Moreover, the middle panel in Figure 8 also shows another obvious absorption at around $-46$ km s$^{-1}$, verifying the spatial consistency between bright radio emission and H I absorption. Thus, the SNR should be located at a distance not nearer than 3.1 kpc so as to explain the absorption at ∼−46 km s$^{-1}$. Furthermore, Figure 8 shows no obvious absorption in the velocity of the tangent point ∼−120 km s$^{-1}$, which gives an upper limit of 7.6 kpc and means the SNR should be located at a near distance. We hence conclude that the SNR is located at a distance $d \sim 3.1$ kpc. This near distance is further supported by the H I self-absorption analysis toward "Region G" (see Appendix).

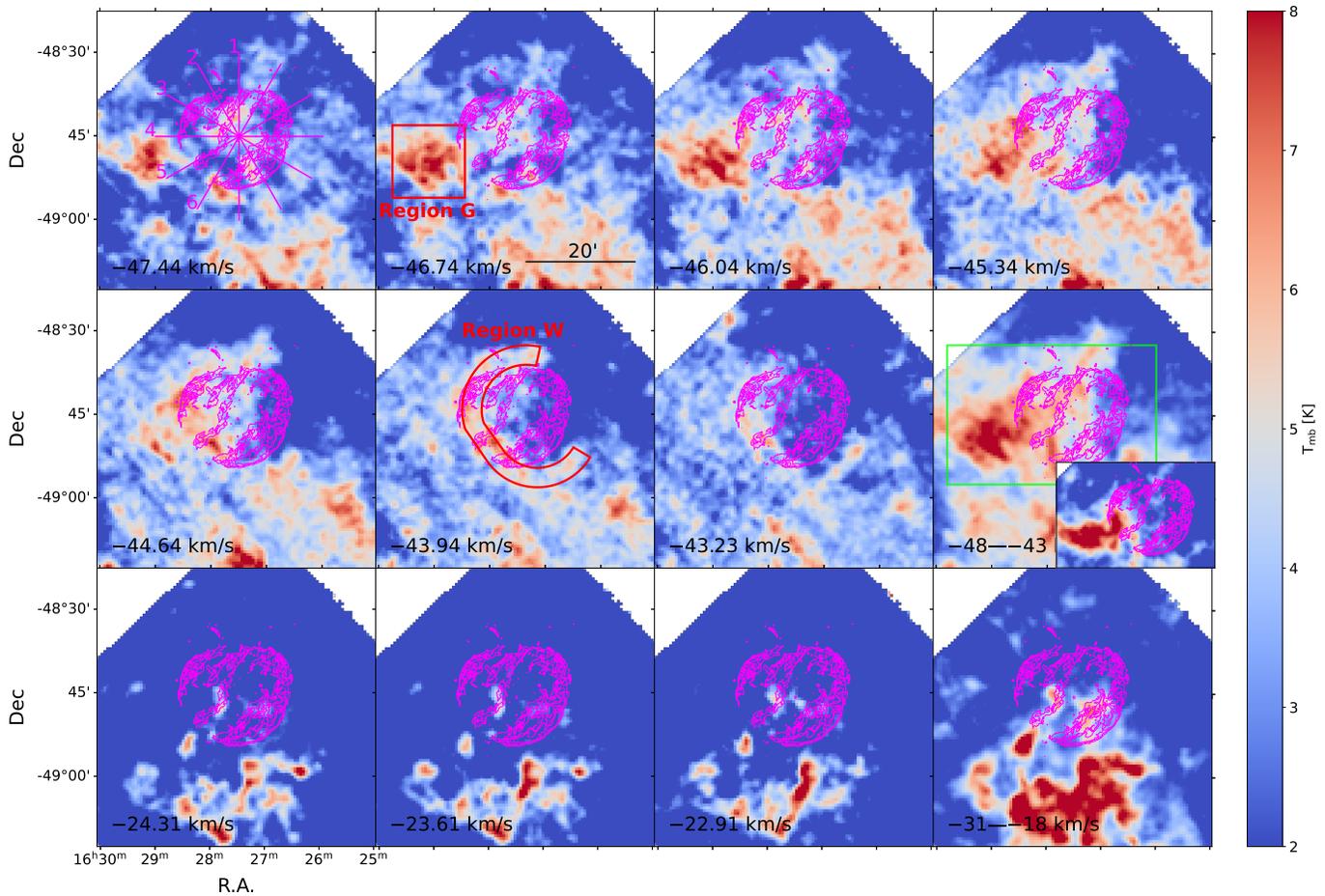

**Figure 5.** Channel maps of the main-beam brightness temperature of $^{12}$CO ($J = 1$–0) emission with a step of 0.7 km s$^{-1}$. The red box, named "Region G," marks the molecular clump coincident with the extended GeV gamma-ray emission. The shell, named "Region W," is the region used to estimate the parameters of the western part of the molecular cloud. The magenta lines passing through SNR G335.2+0.1 are used to extract position–velocity diagrams. The last panel on the second and third rows shows the intensity map of $^{12}$CO emission integrated from $-48$ to $-43$ km s$^{-1}$ (with the $^{13}$CO intensity diagram toward the greenly delineated rectangle region shown in the inset) and from $-31$ to $-18$ km s$^{-1}$, respectively.

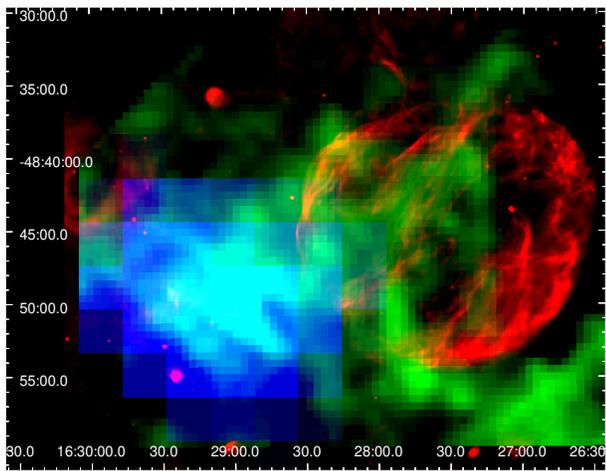

**Figure 6.** Multiwavelength emission map of SNR G335.2+0.1. Red: MeerKAT 1359.7 MHz radio emission (S. Goedhart et al. 2024). Green: intensity map of $^{12}$CO ($J = 1$–0) emission integrated from $-48$ to $-43$ km s$^{-1}$ (detailed in Section 3.2). Blue: the TS map is the same as that in the right panel in Figure 1.

### 4.2. Wind-driven Bubble of the Progenitor Star

The PV diagrams in Section 3.2.2 show an elliptical structure with a velocity width of ∼10 km s$^{-1}$ and a major axis similar to the diameter of the SNR, which is indicative of an expansion motion of the molecular shell at an average velocity $v_b \sim 5$ km s$^{-1}$. A number of similar molecular shells/bubbles expanding with velocities around 5 km s$^{-1}$ have been found surrounding SNRs, such as Tycho's SNR (P. Zhou et al. 2016), VRO 42.05.01 (M. Arias et al. 2019), G352.7−0.1 (Q.-Q. Zhang et al. 2023), G9.7−0.0 (T.-Y. Tu et al. 2025), Kes 67 (Y.-Z. Shen et al. 2025), etc. As in all of these cases, we ascribe the expanding shell in G335.2+0.1 to the bubble in molecular gas driven by the SNR progenitor's wind. On the other hand, for this SNR, if the expanding shell corresponds to the SNR shock, the shock at a velocity as low as ∼5 km s$^{-1}$ would not be able to accelerate electrons to produce the radio synchrotron continuum.

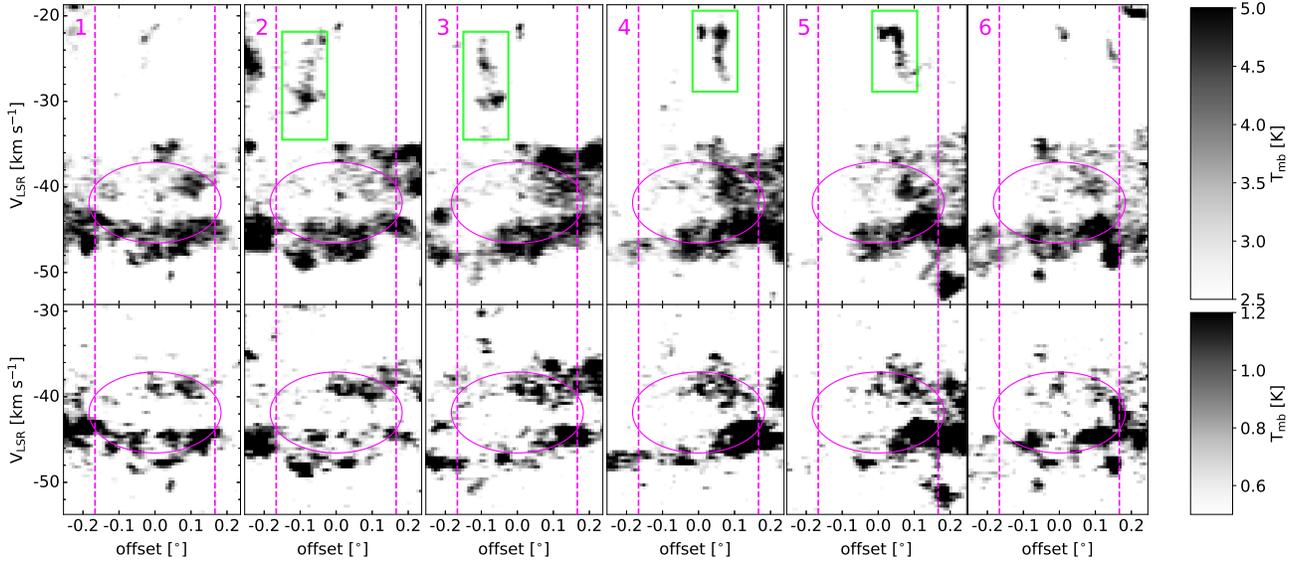

**Figure 7.** Position–velocity diagrams of $^{12}$CO (top) and $^{13}$CO (bottom) emission. The magenta ellipses delineate the expanding structure. The boundary of G335.2+0.1 is marked in a vertical dashed line. The green rectangles denote the possible broadened structure.

**Table 3**
Properties of the Molecular Gas in "Region G" and "Region W"

| Gas Region | $N(H_2)$ ($10^{21}$ cm$^{-2}$) | $M$ ($10^4 M_\odot$) | $n(H_2)$ (cm$^{-3}$) | $T_{ex}$[a] (K) | $\tau(^{13}CO)$[b] |
|---|---|---|---|---|---|
| G | 4.3 | 1.3 | 120 | 15.5 | 0.62 |
| W | 1.4 | 0.4 | 50 | 14.5 | 0.65 |

**Notes.**
[a] Excitation temperature of $^{12}$CO.
[b] Optical depth of $^{13}$CO: $\tau(^{13}CO) \approx -\ln[1 - T_{peak}(^{13}CO)/T_{peak}(^{12}CO)]$.

Using the expansion velocity $v_b \sim 5$ km s$^{-1}$ and adopting the bubble radius $R_b$ as 10.'5 or 9.5 pc (at a distance of 3.1 kpc, see Section 4.1), the timescale of the bubble is estimated by (R. Weaver et al. 1977)

$$t_b = \frac{3R_b}{5v_b} \sim 1 \times 10^6 \left(\frac{R_b}{9.5\,pc}\right)\left(\frac{v_b}{5\,km\,s^{-1}}\right)^{-1} yr. \quad (1)$$

The kinetic luminosity of the progenitor's wind is estimated using the formula (5) in R. Weaver et al. (1977):

$$L_w \sim 4 \times 10^{35} \left(\frac{R_b}{9.5\,pc}\right)^2 \left(\frac{v_b}{5\,km\,s^{-1}}\right)^3 \\ \times \left[\frac{n(H_2)}{50\,cm^{-3}}\right] erg\,s^{-1}, \quad (2)$$

where the molecule density derived from the "Region W" region (see Table 3) is used as a reference value.

If the progenitor is a massive star (other than a single-white-dwarf binary system), according to the linear relation between the molecular bubble's radius and the progenitor's mass (Equation (8) in Y. Chen et al. 2013), the bubble radius $\sim$9.5 pc would imply a progenitor of initial mass $\sim 15\,M_\odot$. This estimate could be regarded as a lower limit if there was energy leakage from the incomplete molecular shell of the bubble.

### 4.3. Evolutionary Status of the Supernova Remnant

Due to the existence of the progenitor's wind bubble, SNR G335.2+0.1 may have first evolved in the cavity inside the bubble before the blast wave struck the cavity wall or the bubble shell. So far, no X-ray emission has been reported at the location of the SNR; also, we have checked the archival XMM-Newton and eROSITA data, and no enhanced X-ray emission was found toward the SNR. So, we consider that the SNR is now in the radiative phase, with the blast shock velocity $v_s$ below 200 km s$^{-1}$, corresponding to the postshock gas temperature below $6 \times 10^5$ K (S. I. Blinnikov et al. 1982). On the other hand, the vivid radio synchrotron continuum emission indicates that electrons are accelerated to relativistic energies. Hence, a lower limit can be set to the shock velocity (B. T. Draine & C. F. McKee 1993): $v_s > 64(x_i/10^{-5})^{1/8}$ $[n(H_2)/50\,cm^{-3}]^{1/8} (\phi_{cr}/0.1)^{-1/4}(T/100\,K)^{0.1}$ km s$^{-1}$, where $x_i$ is the ionization fraction, $\phi_{cr}$ is the efficiency of the particle acceleration of the shock, and $T$ is the preshock molecular gas temperature, but it can be seen that the velocity value is not sensitive to the these quantities. With the higher and lower limits, the shock velocity should be of order $\sim$100 km s$^{-1}$.

According to Figure 5, which shows that the SNR boundary appears to be in contact with the surrounding molecular shell, and the slow shock inferred from lack of X-rays, we consider that the SNR spent its Sedov phase within the wind bubble with a tenuous gas and was switched into the radiative phase upon striking cavity wall (at radius $R_c$) because of drastic deceleration of the shock. This scenario of the SNR evolution is estimated below based on the thin-shell model provided in Y. Chen et al. (2003).

The velocity of the SNR after it stroke the cavity wall could be estimated by

$$v_s = \left(\frac{3E_0 R_c^2}{1.4\pi m_p n_0 R_{sh}^5}\right)^{1/2} F_v^{(R)}(\lambda_c), \quad (3)$$

where $\lambda_c = R_c/R_{sh}$ (that is, the ratio between the transition radius from adiabatic phase to radiative phase and the current SNR radius), $E_0$ is the explosion energy (assumed to be 10$^{51}$ erg here), $m_p$ is the proton mass, and $n_0$ is the number density of the preshock hydrogen nuclei. The function $F_v^{(R)}(\lambda_c)$ is given by

$$F_v^{(R)}(\lambda_c) = \left\{\left(1 - \lambda_c\right) + \frac{\mu}{2}\lambda_c(\lambda_c^2 - 1)\right. \\ \left. + \frac{\lambda_c}{3}[1 - \mu(1 + 3\ln\eta)]\right\}^{1/2}(1 - \mu\lambda_c^3)^{-1}, \quad (4)$$

where $\eta$ is the ratio between the transition radius and the cavity wall radius (approximated as unity here), and $\mu = (1 - \mathcal{B})/\eta^3$, with $\mathcal{B} = n_1/n_0$ the density contrast between the bubble interior and the cavity wall. In search of appropriate parameters, $\mathcal{B}$ = 0.0005 and 0.001 are used, for example.

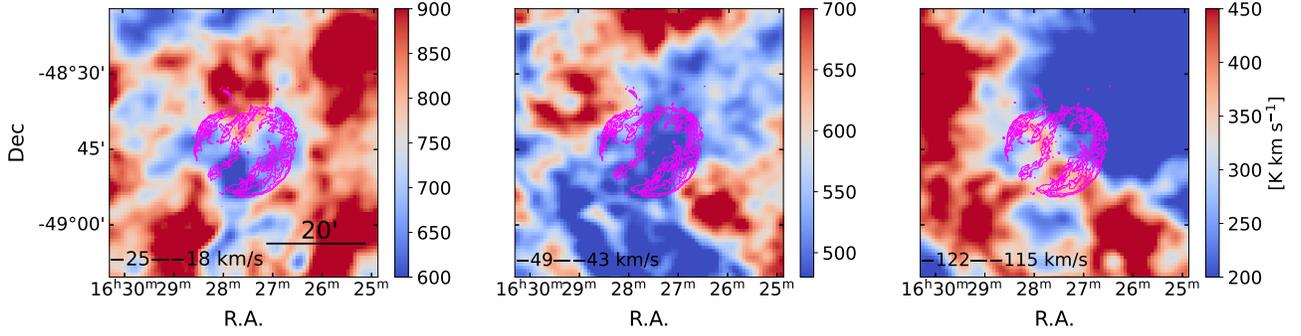

**Figure 8.** Intensity maps of SGPS H I emission in the velocity range −25 to −18 km s$^{-1}$, −49 to −43 km s$^{-1}$, and −122 to −115 km s$^{-1}$ with magenta contours the same as those in Figure 1.

The SNR age $t$ is obtained by

$$t - t_c = [F_r^{(R)}(\lambda_c)]R_{sh}^{7/2}\left(\frac{3E_0R_c^2}{1.4\pi m_p n_0}\right)^{-1/2}, \quad (5)$$

where $t_c$ is the age when the SNR evolved into the radiative phase and is estimated using the L. I. Sedov (1959) self-similar solutions. The function $F_r^{(R)}(\lambda_c)$ is

$$F_r^{(R)}(\lambda_c) = \int_{\lambda_c}^{1} \frac{\lambda^{5/2}}{F_v^{(R)}(\lambda_c/\lambda)} d\lambda. \quad (6)$$

The numerical values of $v_s$ and $t$ are plotted in Figure 9. The gas density inside the bubble, $n_1$, should be no smaller than 0.1 cm$^{-3}$, derived assuming that about 10 $M_\odot$ (from an initial mass of ∼15 $M_\odot$ progenitor star, see Section 4.2) was ejected into the wind-blown bubble. However, $n_1$ could exceed this lower limit due to mass loading processes. For example, small dense clumps or cloudlets engulfed in the bubble could contribute mass via UV photoevaporation and wind destruction. Moreover, the swept-up ambient molecular gas evaporated into the shocked wind region by thermal conduction might even dominate the interior mass (J. Castor et al. 1975). We thus adopt 0.5 and 1 cm$^{-3}$ as reference density values in the model calculation. The SNR age is then estimated to be $t$ ∼ 3.6 kyr and ∼5.1 kyr for $\mathcal{B}$ = 0.0005 and 0.001, respectively. In the following discussion, we adopt the combination of parameters ($n_1$, $\mathcal{B}$, $t$, $t_c$) = (1 cm$^{-3}$, 0.001, 5.1 kyr, 5.0 kyr) as exemplified parameters in the following discussion.

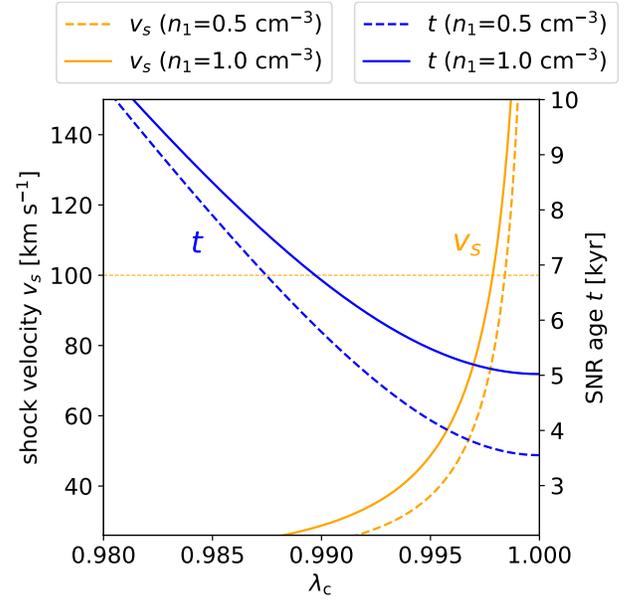

**Figure 9.** Plots of the modeled shock velocity $v_s$ (orange lines) and SNR age $t$ (blue lines) vs. parameter $\lambda_c$. The dashed and solid curves are plotted for $\mathcal{B}$ = 0.0005 and 0.001, respectively. The horizontal dashed line denotes $v_s$ = 100 km s$^{-1}$.

### 4.4. Origin of Gamma-Ray Emission Associated with SNR G335.2+0.1

The analysis of Fermi-LAT data in Section 3.1 reveals an extended gamma-ray emission partially overlapping with SNR G335.2+0.1. Except for the SNR as a high-energy source, no known pulsars are found projectively within the disk model radius according to the Australia Telescope National Facility pulsar catalog[9] (R. N. Manchester et al. 2005), and the nearest pulsar PSR J1627−4845 is 0°.307 away from the center of the GeV source. This pulsar is located at a distance of 4.9 kpc with a relatively low spin-down power, 6.3 × 10$^{32}$ erg s$^{-1}$, that cannot afford the GeV gamma-ray flux ∼1.4 × 10$^{-11}$ erg s$^{-1}$ cm$^{-2}$. In addition, a high mass X-ray binary IGR J16283−4838 is 0°.20 away from the center of the GeV source, which is far beyond the 99% position uncertainty of the disk (0°.08). Thus we suggest that it is unrelated to the GeV source.

The revealed GeV gamma rays associated with SNR G335.2 +0.1 are soft and luminous, with an index 2.0–2.2 (see Table 1) and a large luminosity (∼2 × 10$^{34}$ erg s$^{-1}$,

---
[9] https://www.atnf.csiro.au/research/pulsar/psrcat/

Section 3.1.2), appearing consistent with a hadronic origin. SNR-associated GeV gamma rays have been found to usually have photon indices above 2 and luminosities above $10^{34}$ erg s$^{-1}$ (see, e.g., B. Liu et al 2015 and Table 3 therein; F. Acero et al. 2016 and Table 2 therein; W.-J. Zhong et al. 2023).

The extended GeV gamma-ray source almost overlaps the MC at $\sim -46$ km s$^{-1}$ which very likely interacts with the SNR G335.2+0.1. The gamma-ray luminosity of the bremsstrahlung process is compatible with that of the hadronic process if the number ratio of electrons to protons is of the order of $\sim 0.1$ (T. K. Gaisser et al. 1998). However, the number ratio predicted by diffusive shock acceleration theory (A. R. Bell 1978) is $\sim 0.01$. Thus the bremsstrahlung gamma ray is ignored here. Naturally, the gamma-ray emission is likely to arise from the molecular gas bombarded directly by the protons trapped in the SNR or illuminated by the escaped protons diffusing from the SNR shock. Considering that the GeV source is mainly outside the radio extent of the SNR, we only considered the case of illumination by the escaped protons.

In the following model calculation of the gamma-ray SED, the shape of the large clump ("Region G") is approximated as a truncated cone that subtends a solid angle at the SNR center (as treated similarly in H. Li & Y. Chen 2012), with the inner and outer radii adopted as the distances of the near edge and far edge from the SNR center, $L_1$ and $L_2$ (see Section 3.2.3), respectively. According to Sections 4.1 and 4.3, the distance to SNR G335.2+0.1 is $d \sim 3.1$ kpc, the radius of SNR is $\sim 9.5$ pc, and an age $\sim 5.1$ kyr is adopted. We also assume the explosion energy $E_0 = 10^{51}$ erg, the initial ejecta velocity $V_0 = 5 \times 10^3$ km s$^{-1}$, and adopt the gas density in the progenitor's wind cavity $n_1 \sim 1$ cm$^{-3}$. The SNR radius was $R_{\rm Sedov} \sim 3.0$ pc when it entered the Sedov phase, i.e., when the mass of the swept-up medium was compatible with the ejecta mass.

We first tested the illumination model developed in H. Li & Y. Chen (2012), in which the accelerated protons, including those below the maximum energy, are considered to escape from the expanding SNR during its evolution. This may mimic the case in which the shock propagates in a clumpy medium, and energetic particles may leak from the broken portions of the shell and diffuse to the MC. The protons are assumed to escape uniformly with time in both the Sedov phase and the radiative phase. The distribution of escaping protons is momentum (energy) dependent $\propto p^{-s_{\rm esc}}$, where $s_{\rm esc}$ is the CR index after they escape from the SNR. The momentum-dependent diffusion coefficient is

$$D(p) = 10^{28} \chi \left(\frac{cp}{10 \text{ GeV}}\right)^\delta \text{cm}^2\text{s}^{-1}. \quad (7)$$

On the assumption of a spherically symmetric system, CR particles leaving the shock will diffuse a distance

$$R_{\rm d}(p) = \sqrt{4D(p)(t - t_{\rm esc})}, \quad (8)$$

where $t_{\rm esc}$ is the time when CR particles leave the shock. The mean distribution of the energetic protons in the target MC (between $L_1$ and $L_2$) is calculated using equation (2) in H. Li & Y. Chen (2012). The maximum energy of particles in the SNR cannot be constrained by the Fermi data and, hence, is fixed to be $10^{15}$ eV. As can be seen in Figure 2, this simple model (the blue line) can fit the GeV gamma-ray SED well with relevant parameters listed in Table 4.

**Table 4**
Parameters Used in the Diffusion Model

| Parameters | Value | Reference |
|---|---|---|
| $d$ (kpc) | 3.1 | this work |
| $t_{\rm age}$ (kyr) | 5.0 | this work |
| $L_1$ (pc) | 11 | fixed |
| $L_2$ (pc) | 20 | fixed |
| $cp_{\rm knee}$ (eV) | $10^{15}$ | fixed |
| H. Li & Y. Chen (2012) | | |
| $M_{\rm c}$ ($10^4 M_\odot$) | 0.5 | fitted |
| $\chi$ | 0.1 | fitted |
| $\delta$ | 0.6 | fitted |
| $s_{\rm esc}$ | 2.1 | fitted |
| Y. Ohira et al. (2011) | | |
| $\kappa$ | 0.1 | (1) |
| $M_{\rm c}$ ($10^4 M_\odot$) | 0.6 | fitted |
| $\alpha$ | 7.5 | fitted |
| $\chi$ | 0.2 | fitted |
| $\delta$ | 0.6 | fitted |
| $\beta$ | 1.7 | fitted |
| $s$ | 1.9 | fitted |

**Reference.** (1) V. S. Ptuskin & V. N. Zirakashvili (2005).

In an alternative model, we adopted the so-called $\delta$-escape process, in which, at a certain time, only the protons with the maximum energy can escape at the free-escape boundary and then diffuse into the ambient medium (e.g., V. S. Ptuskin & V. N. Zirakashvili 2005). The SNR radius $R_{\rm esc}$ at which the protons escape was derived by adopting a phenomenological approach based on the power-law dependence (S. Gabici et al. 2009; Y. Ohira et al. 2011)

$$R_{\rm esc}(p) = (1 + \kappa) R_{\rm Sedov} \left(\frac{p}{p_{\rm knee}}\right)^{-1/\alpha}, \quad (9)$$

where $\kappa$ is the fractional thickness of the precursor ahead of the SNR shock confining the accelerated particles before escape, adopted as 0.1 in terms of a high-velocity regime (V. S. Ptuskin & V. N. Zirakashvili 2005), $cp_{\rm knee} = 10^{15}$ eV is the highest energy with which the particle can escape from the SNR at the start of the Sedov phase, and $\alpha$ is a variable within a reasonable interval. Note that once the shock collides with the bubble wall (that is, $p = p_{\rm bre}$ when $R_{\rm esc}(p) = R_{\rm c}$), the particles with $p \leqslant p_{\rm bre}$ escape almost at the same time (Y. Ohira et al. 2011). The SNR age $t_{\rm esc}$ when $R_{\rm sh} = R_{\rm esc}(p)/(1 + \kappa)$ is obtained using the Sedov self-similar solutions (L. I. Sedov 1959). Thus, the maximum distance the CRs with momentum $p$ can reach is $R_{\rm esc}(p) + R_{\rm d}(p)$.

Moreover, we also assume that the number of CRs in the SNR is $K(R_{\rm sh})dp \propto R_{\rm sh}^\beta$ and the CR spectrum accelerated at the shock front is $\propto p^{-s}$. Thus, the simplest form of the escaped CRs spectrum is $N_{\rm esc}(p) \propto p^{-(s + \beta/\alpha)}$ (Y. Ohira et al. 2010).

We applied the distribution of the runaway CRs (see equation (6) in Y. Ohira et al. 2011) to reproduce the observed gamma-ray spectrum. From the model, the CR distribution function in the region between $L_1$ and $L_2$ is obtained. The fitting model curve is also shown in Figure 2, and the gamma-ray SED is reproduced as well. The fitted parameters are similar to those for the H. Li & Y. Chen (2012) model (see Table 4). Taking the particle indices as an example, given the index $s = 1.9$, we also obtained a CR spectral index $s_{\rm esc} = 2.1$ according to the thermal leakage model $\beta = 3(3 - s)/2$ (Y. Ohira et al. 2010). This value has been well constrained (A. W. Strong & I. V. Moskalenko 1998) and can reproduce the observed Galactic CR spectrum on Earth with the propagation effect taken into account. The fitted gas mass, $0.5$–$0.6 \times 10^4 M_\odot$, is similar for the two models. They are somewhat smaller than the total derived mass of molecular gas in the "Region G" and "Region W," implying either that part of the gas in the region is not subject to the hadronic process or the derived gas mass is overestimated.

Both of the escaped particle illumination models adopted, in which the effect caused by the finite volumes of the source and the MC was considered, account for the spectral breaks (H. Li & Y. Chen 2010; Y. Ohira et al. 2011; H. Li & Y. Chen 2012). The resulting index of diffusion coefficient was constrained to be $\delta = 0.6$, close to the value required by CR propagation models (V. S. Berezinskii et al. 1990), while the diffusion constant $\chi = 0.1$ is an order lower than the Galactic mean or that of the classic CR propagation models. A possible explanation for such a small value is that the runaway CRs amplify the magnetic turbulence, accounting for a significant suppression of the diffusion process. Actually, the suppressed diffusion was also observed in SNR W28 (Y. Fujita et al. 2009; H. Li & Y. Chen 2012; Y. Hanabata et al. 2014) and W44 (Y. Uchiyama et al. 2012; S. Abe et al. 2025).

By the above fit of SED with the two models, we conclude that the observed gamma-ray emission can be naturally interpreted with a hadronic process between the CRs escaped from SNR G335.2+0.1 and the nearby molecular clump ("Region G"). Hence, SNR G335.2+0.1 unambiguously joins the list of confirmed cases of CR-illuminated MCs associated with SNRs, such as W28 (F. Aharonian et al. 2008b; H. Li & Y. Chen 2010; Y. Hanabata et al. 2014), W44 (Y. Uchiyama et al. 2012; S. Abe et al. 2025), and HB 9 (T. Oka & W. Ishizaki 2022; Y. Bao et al. 2024).

But the diffusion models are based on some simplified assumptions. The diffusion coefficient of CRs was regarded as spatially uniform and independent of CR density. From previous studies, the escaped CRs can disturb the surrounding magnetic field, which may lead to the suppression of diffusion (R. Kulsrud & W. P. Pearce 1969; D. G. Wentzel 1969). Moreover, we simply assumed a time-independent particle distribution in the shock during the total process, but the fact is more complicated.

In the TeV band, the extended gamma-ray source HESS J1626−490 is $0\overset{\circ}{.}42$ away from SNR G335.2+0.1. We checked the $^{12}$CO and $^{13}$CO data but no dense molecular clump was found coincident with the TeV gamma-ray source at $\sim -48$ to $-43 \rm\ km\ s^{-1}$. There is indeed a partial overlap along the line of sight between HESS J1626−490 and the MCs at $-31$ to $-25 \rm\ km\ s^{-1}$; however, there is no evidence of association between these MCs and the SNR.

## 5. Conclusion

The possible correlation between TeV source HESS J1626−490 and SNR G335.2+0.1 suggested by previous research has motivated us to study the GeV gamma-ray emission associated with the SNR and the molecular environment of the SNR. Using the 16.8 yr Fermi-LAT data at the location of G335.2+0.1, we find an extended GeV emission, with a significance of $13.5\sigma$ in 0.2–500 GeV, which is located to the east of the SNR and partially overlaps with it. We do not find a GeV counterpart for HESS J1626−490.

To explore the origin of the extended GeV gamma-ray emission, we investigated the molecular environment of the SNR, employing the $^{12}$CO and $^{13}$CO data from Mopra and H I data from SGPS. We found that the SNR is located in a cavity encircled by a "C"-shaped ringlike incomplete molecular shell at $V_{\rm LSR} \sim -45$ to $-43 \rm\ km\ s^{-1}$. A large molecular clump ("Region G"), to the east of the SNR at $V_{\rm LSR} \sim -48$ to $-43 \rm\ km\ s^{-1}$ appears coincident with the GeV gamma-ray emission.

The PV diagrams, which are produced along lines crossing through the geometric center of the SNR at different positional angles, display arclike or elliptical structures across the remnant between $-47$ and $-37 \rm\ km\ s^{-1}$, indicating an expansion motion (at $\sim$5 km s$^{-1}$) of the surrounding molecular gas. This motion, along with the molecular cavity, is ascribed to the role of the stellar wind of the SNR progenitor. On the basis of the morphological agreement of the molecular cavity with the SNR, the PV diagrams provide kinematic evidence of the association of the MCs around $-46 \rm\ km\ s^{-1}$ are physically associated with the SNR G335.2+0.1. This LSR velocity, together with the inspected H I absorption, places the SNR at a distance of 3.1 kpc. The initial mass of the progenitor is estimated to be $\gtrsim 15\ M_\odot$.

We suggest an evolutionary scenario in which the SNR shock has recently struck the cavity wall after a $\sim$5 kyr expansion within the wind-blown bubble, suffering a drastic deceleration, and entered the radiative phase.

Incorporating this evolutionary scenario of SNR G335.2 +0.1, we demonstrated that the extended GeV gamma-ray emission may arise from the eastern large molecular clump, considering that the clump is bombarded by the energetic protons that escape from the SNR shock. The observed SEDs are well fitted by applying two hadronic illumination models developed by H. Li & Y. Chen (2012) and Y. Ohira et al. (2011), respectively. The fitted index $\approx 2.1$ is consistent with the value predicted by the classical diffusive shock acceleration. The fitted diffusion coefficient is an order of magnitude lower than the Galactic mean or that of the classic CR propagation models, possibly due to the magnetic turbulence.

## Acknowledgments

We thank Yun-Zhi Shen, Tian-Yu Tu, and S. Ranasinghe for useful advice. This study is supported by NSFC under grants 12173018, 12121003, and 12393852.

*Facilities:* Fermi, Mopra, MeerKAT, ATCA, Parkes.

*Software:* Fermitools, Fermipy (M. Wood et al. 2017), Naima (V. Zabalza 2015), Astropy (Astropy Collaboration et al. 2022).

## Note Added in Manuscript

After the submission of this paper, we became aware of a preprint by T. Oka et al. (2025), which reports similar results on the GeV gamma-ray emission and the molecular environment associated with SNR G335.2+0.1. But there are notable differences between the two works. Our work emphasizes that the SNR evolved in an expanding molecular bubble blown by the progenitor's wind and suggests that the SNR shock has entered the radiative phase after striking the cavity wall. Furthermore, we interpret that most of the GeV gamma-ray emission originates from the large eastern molecular clump illuminated by the diffusive protons that have escaped from the shock front.

## Appendix
## H I Self-absorption toward "Region G"

For better determining the distance to the molecular gas of interest at around $-46\,\mathrm{km\,s^{-1}}$ and the associated SNR independently, we subsidiarily adopted the H I self-absorption (HISA) method (see, e.g., J. Roman-Duval et al. 2009). We examined the HISA toward the on-source region (i.e., "Region G") by comparing the H I spectrum from the on-source region with that from the off-source region (marked "off" in Figure A1). The off-source region lies within $0.\!\!^\circ2$ of the center of "Region G." As can be seen in Figure A1, the on-source region (with a $^{13}$CO emission peak) exhibits an H I absorption line at the same LSR velocity as that ($-45\,\mathrm{km\,s^{-1}}$) of the $^{13}$CO peak, whereas no H I absorption appears in the off-source region. This indicates that the molecular gas of interest lies at a near kinematic distance.

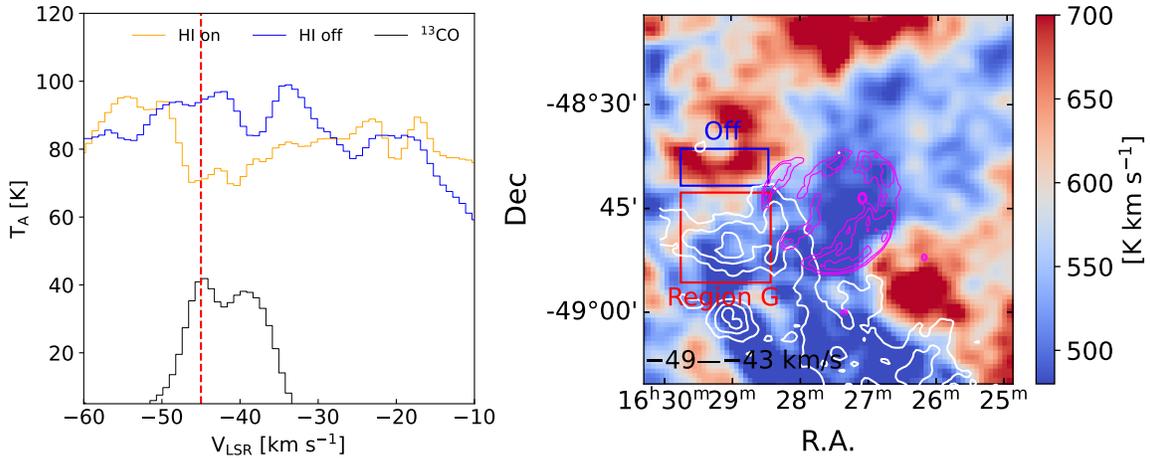

**Figure A1.** Illustration of H I self-absorption toward "Region G." Left panel: H I spectra toward the on-source region ("Region G") and off-source position (as marked "off" in the right panel) for the molecular gas in "Region G," together with the $^{13}$CO spectrum from "region G" (scaled by 50 times for comparison with H I spectra). The vertical line denotes the peak LSR velocity $\sim -45\,\mathrm{km\,s^{-1}}$ of the absorption. Right panel: the same SGPS H I emission intensity map as in the middle panel of Figure 8, but with the intensity contours (in white) of $^{13}$CO emission in the same velocity range. The red box and blue rectangle show the on-source region ("region G") and the off-source region (marked "off"), respectively.

## ORCID iDs

Chen Huang ⬥ https://orcid.org/0000-0003-4916-4447
Xiao Zhang ⬥ https://orcid.org/0000-0002-9392-547X
Yang Chen ⬥ https://orcid.org/0000-0002-4753-2798
Qian-Qian Zhang ⬥ https://orcid.org/0000-0003-0853-1108
Wen-Juan Zhong ⬥ https://orcid.org/0000-0003-3717-2861
Xin Zhou ⬥ https://orcid.org/0000-0003-2418-3350